\documentclass[prl,twocolumn,amsmath,amssymb]{revtex4}

\usepackage{amsmath}
\usepackage{amssymb}
\usepackage{float}
\usepackage{graphicx}
\usepackage{upgreek}

\usepackage{color}

\makeatletter

\newcommand{\Rmnum}[1]{\expandafter\@slowromancap\romannumeral
#1@}
\makeatother

\usepackage{float}
\usepackage[T1]{fontenc}
\usepackage{indentfirst}
\begin{document}

\title{N\'{e}el and stripe ordering from spin-orbital entanglement in $\alpha$-Sr$_2$CrO$_4$}
\author{Z. H. Zhu$^1$}\email{zzh@mit.edu}

\author{W. Hu$^2$}
\author{C. A. Occhialini$^1$}
\author{J. Li$^1$}
\author{J. Pelliciari$^1$}
\author{C. S. Nelson$^3$}
\author{M. R. Norman$^4$}
\author{Q. Si$^2$}

\author{R. Comin$^1$}
\email{rcomin@mit.edu}  
\affiliation{$^1$Department of Physics, Massachusetts Institute of Technology, Cambridge, MA 02139, USA}
\affiliation{$^2$Department of Physics and Astronomy, Rice Center for Quantum Materials, Rice University, Houston, TX 77005, USA}
\affiliation{$^3$National Synchrotron Light Source II, Brookhaven National Laboratory, Upton, NY 11973, USA}
\affiliation{$^4$Materials Science Division, Argonne National Laboratory, Argonne, IL 60439, USA}

\date{\today}

\begin{abstract}
The rich phenomenology engendered by the coupling between the spin and orbital degrees of freedom has become appreciated as a key feature of many strongly-correlated electron systems.  The resulting emergent physics is particularly prominent in a number of materials, from Fe-based unconventional superconductors to transition metal oxides, including manganites and vanadates.  Here, we investigate the electronic ground states of $\alpha$-Sr$_2$CrO$_4$, a compound that is a rare embodiment of the spin-1 Kugel-Khomskii model on the square lattice -- a paradigmatic platform to capture the physics of coupled magnetic and orbital electronic orders.  We have used resonant X-ray diffraction at the Cr-$K$ edge to reveal N\'{e}el magnetic order at the in-plane wavevector $\mathbf{Q}_N = (1/2, 1/2)$ below $T_N = 112$\,K, as well as an additional electronic order at the ``stripe'' wavevector $\mathbf{Q}_s = (1/2, 0)$ below T$_s$ $ \sim 50$\,K.  These findings are examined within the framework of the Kugel-Khomskii model by a combination of mean-field and Monte-Carlo approaches, which supports the stability of the spin N\'{e}el phase with subsequent lower-temperature stripe orbital ordering, revealing a candidate mechanism for the experimentally observed peak at $\mathbf{Q}_s$. On the basis of these findings, we propose that $\alpha$-Sr$_2$CrO$_4$ serves as a new platform in which to investigate multi-orbital physics and its role in the low-temperature phases of Mott insulators.
\end{abstract}

\maketitle
Multi-orbital physics plays an important role for several collective electronic phenomena in strongly-correlated systems, including high-$T_c$ superconductivity and exotic magnetism \citep{streltsov_orbital_2017,tokura_orbital_2000}. Some model materials, such as LaMnO$_3$ \citep{murakami_resonant_1998}, KCuF$_3$ \cite{caciuffo_resonant_2002}, and YVO$_3$ \citep{noguchi_synchrotron_2000}, have enabled the exploration of the active role played by the orbital degrees of freedom in determining the symmetry of the electronic ground state.  In particular, the essential ingredients of unconventional superconductivity in the iron-based superconductors (FeSCs) are purported to involve multi-orbital electronic states and their interplay with magnetism \citep{yi_role_2017,liu_0_2010,stewart_superconductivity_2011}, in remarkable contrast to the effective single-orbital nature of cuprate superconductors \citep{lee_doping_2006}.

A few chromate perovskites ACrO$_3$ (A =Ca, Sr, and Pb) with a Cr$^{4+}$ valence state were synthesized by several groups four decades ago \citep{chamberland_preparation_1967,chamberland_study_1972,goodenough_band_1968}. Recent studies on CaCrO$_3$ and SrCrO$_3$ have revealed novel physics, drawing more attention to these chromate compounds \citep{zhou_anomalous_2006,ortega-san-martin_microstrain_2007,komarek_2008}. For example, both compounds are antiferromagnetic conductors, but the exact nature of the ground state has remained elusive.  An interesting orbital ordering transition with electronic phase coexistence has been discovered in SrCrO$_3$ \citep{ortega-san-martin_microstrain_2007}, though it is unclear if the orbital degree of freedom is essential to carrier itinerancy within the antiferromagnetic ground state. Moreover, despite the isovalency of Ba and Sr, the sister compound BaCrO$_3$ was found to be an antiferromagnetic insulator \citep{zhu_magnetic_2013,arevalo-lopez_high-pressure_2015}, with theoretical studies proposing that the mechanism for the insulating state involves, again, some form of orbital ordering \citep{giovannetti_cooperative_2014,jin_strain_2014}. A recent study of bilayered Sr$_3$Cr$_2$O$_7$ revealed the presence of an exotic ordered phase of orbital singlets arising from the interplay between the spin and orbital degrees of freedom \citep{jeanneau_singlet_2017}.  It is thus not surprising that similar phenomena at the nexus between magnetism and orbital physics were proposed also for the single-layer ($n=1$) member of the Ruddlesden-Popper series of chromates, Sr$_2$CrO$_4$ \citep{ishikawa_reversed_2017}. However, the microscopic nature of the ground state of this compound remains unknown.

In the present work, we focus on $\alpha$-Sr$_2$CrO$_4$, which crystallizes in the K$_2$NiF$_4$-type structure that makes it isostructural to La$_2$CuO$_4$. With its rare Cr$^{4+}$ oxidation state, it realizes a unique $3d^2$ electronic configuration among layered oxides, resulting in active $t_{2g}$ orbitals with total spin $S=1$ on a square lattice. Despite several studies characterizing this unique material \citep{sakurai_synthesis_2014,nozaki_2018,sugiyama_microscopic_2014}, a clear picture of its physical properties has remained elusive, partly owing to known challenges with the synthesis of bulk single crystals. Successful growth of polycrystalline samples of $\alpha$-Sr$_2$CrO$_4$ by means of high-pressure ($> 5$\,GPa) and high-temperature (1500\,$^{\circ}$C) synthesis has been reported by several groups\citep{sakurai_synthesis_2014,baikie_crystallographic_2007,chamberland_magnetic_1985}. Magnetization characterization of these polycrystalline samples reveal a sharp antiferromagnetic transition at T$_N$ $\approx 112$\,K\cite{sakurai_synthesis_2014}, which was further confirmed by muon spin rotation ($\mu^+$SR) measurements \citep{nozaki_2018,sugiyama_microscopic_2014}. The resistivity of sintered pellets of $\alpha$-Sr$_2$CrO$_4$ reveal insulating behavior \citep{sakurai_synthesis_2014}. In addition, single-crystalline samples of $\alpha$-Sr$_2$CrO$_4$ have been synthesized as thin films using pulsed laser deposition (PLD), which are determined by optical conductivity to be Mott-Hubbard type insulators with a charge gap $\Delta$ $\approx$\,0.3 eV \citep{matsuno_variation_2005}. While previous studies provide evidence for an antiferromagnetic insulating ground state in $\alpha$-Sr$_2$CrO$_4$, no direct reciprocal space studies of the long-range electronic order in these compounds have been reported to date.

\begin{figure*}[ht!]
\centering
\includegraphics[width=1.6\columnwidth]{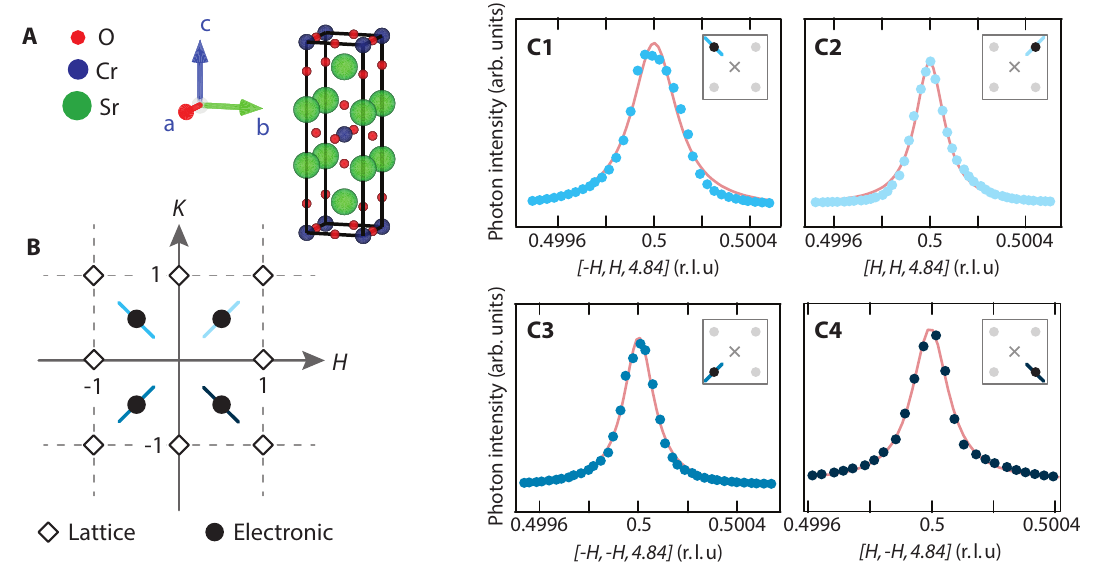}
\caption{\label{fig1} \textbf{Resonant X-ray scattering measurements of $\alpha$-Sr$_2$CrO$_4$ at the Cr-$K$ edge.} \textbf{A}: unit cell of $\alpha$-Sr$_2$CrO$_4$, which is tetragonal with \textit{I4/mmm} space group at room temperature. \textbf{B}: Reciprocal space cartography of scattering scans in the $\left( H, K \right)$ plane. \textbf{C1-C4}: individual RXS scans across the symmetry-equivalent ordering vectors $\left(\pm 0.5, \pm 0.5, 4.84 \right)$ (see insets).}
\end{figure*}

To address this open question, we have synthesized single-crystalline films of $\alpha$-Sr$_2$CrO$_4$ with the aim of resolving its magnetic ground state and to search for other ordered phases within the electronic degrees of freedom.  We have sought reciprocal space signatures of electronic long-range ordering using resonant x-ray scattering (RXS) measurements at the Cr-$K$ edge (with photon energy $E_{ph} \approx 6.0$\,keV). At the Cr-$K$ edge, photons that are tuned to the Cr-$1s \rightarrow 4p$ transition become resonantly sensitive to the charge, spin, and orbital degrees of freedom at the Cr site, a capability that has made RXS a valuable probe to study electronic ordering in transition metal oxides \citep{paolasini_resonant_2014,matteo_resonant_2012}.

Thin film samples of $c$-oriented Sr$_2$CrO$_4$ have been grown on (001) SrTiO$_3$ substrates using pulsed laser deposition (PLD) from a KrF excimer laser ($\lambda = 248$\,nm). Deposition was performed under high vacuum with a laser pulse energy and repetition rate of 400\,mJ and 4\,Hz, respectively, with the substrate heated to 900\,$^{\circ}$C.  The laser target contained a stoichiometric mixture of SrCO$_3$ and Cr$_2$O$_3$ prepared by a solid state reaction at 920\,$^{\circ}$C under a constant N$_2$ flow for 24 hours. The products of this reaction were ground and pressed into a pellet, then sintered under an Ar flow at 1200\,$^{\circ}$C for 24 hours. The main phase of the target was determined to be $\beta$-Sr$_2$CrO$_4$, however the final $\sim 100$nm film was confirmed by X-ray diffraction to be pure $\alpha$-Sr$_2$CrO$_4$, with no traces of the sister phase (see Supplementary Information, in particular Fig.\,S1, for additional details). Resonant X-ray measurements were performed at the Integrated In-Situ and Resonant X-ray Studies (ISR) beamline- 4-ID of NSLS-II at Brookhaven National Laboratory.

\begin{figure}[h]
\centering
\includegraphics[width=1\columnwidth]{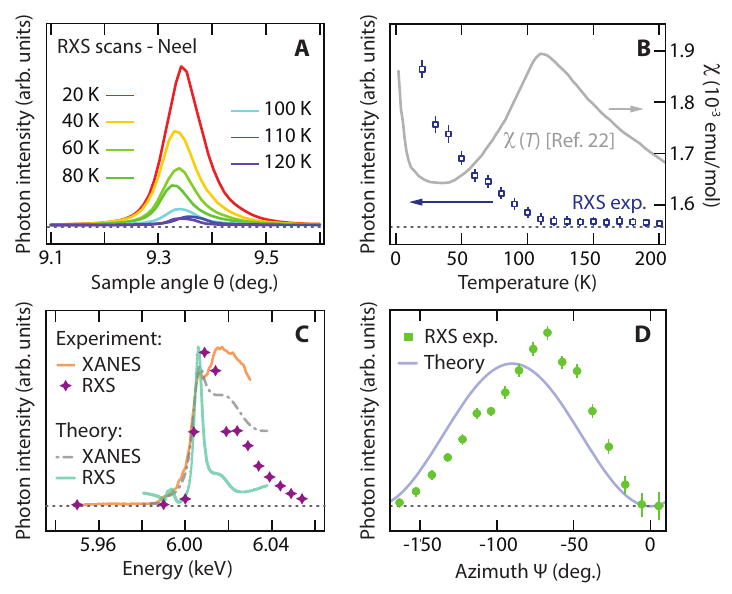}
\caption{\label{fig2} \textbf{Nature and characteristics of the N\'{e}el-ordered state.} \textbf{A}: energy-integrated sample rocking scans across the resonant reflection ${\mathbf{Q}}_{N} = \left( 1/2, 1/2, L \right)$ at various temperatures. \textbf{B}: temperature dependence of the integrated scattering intensity around ${\mathbf{Q}}_{N}$ (markers) and of the magnetic susceptibility $\chi \left( T \right)$, from \citep{sakurai_synthesis_2014} (gray line). \textbf{C}: energy dependence of the integrated RXS intensity around ${\mathbf{Q}}_{N}$ (purple diamonds) and near-edge X-ray absorption scan (XANES, orange trace), both acquired at 20\,K; the dash-dotted gray and teal lines represent, respectively, simulations of the XANES spectrum and the RXS signals from magnetic scattering using FDMNES program (see text and Supplementary Information). \textbf{D}: azimuthal angle ($\psi$) dependence of the integrated intensity (green squares) of the ${\mathbf{Q}}_{N}$ reflection, with the blue curve representing the simulated azimuthal dependence expected for the magnetic model described in the text.}
\vspace{0.0cm}
\end{figure}

Figure\,\ref{fig1}A shows the unit cell of $\alpha$-Sr$_2$CrO$_4$, which is tetragonal with space group \textit{I4/mmm} at room temperature. Throughout this paper, reciprocal lattice vectors \textit{(H,K,L)} are scaled to reciprocal lattice units (r.l.u) for the room temperature tetragonal structure (that is, in units of $2\pi / a$, $2\pi / a$, $2\pi / c$, where $a = b = 3.87$\,\AA, $c = 12.55$\,\AA). Resonant X-ray scattering scans were performed in the reciprocal $\left( H, K \right)$ plane parallel to the crystallographic CrO$_2$ layers, as illustrated in Figure\,\ref{fig1}B. We have surveyed the high-symmetry directions $\left[ H, 0 \right]$ and $\left[ H, H \right]$ using incident X-rays tuned to the Cr-$K$ edge resonance E = 6.009\,keV to resonantly enhance Bragg reflections of electronic nature. Under these experimental conditions, we have observed four well-defined, symmetry-equivalent diffraction peaks at the N\'{e}el vectors ${\mathbf{Q}}_{N} = \left( \pm 1/2, \pm 1/2, L \right)$ (r.l.u), with $L \sim 4.84$, as shown in Fig.\,\ref{fig1}C1-C4.  These diffraction signatures reveal the presence of a $\sqrt{2} \times \sqrt{2}$ electronic superlattice which is forbidden by the high-symmetry structure. The disappearance of the superlattice reflection away from the Cr-$K$ resonance suggests that it results from a reordering of the electronic subspace. The in-plane correlation length $\xi_{ab}$ = a$(\pi FWHM)^{-1}$ $\sim $ 4000 \AA ($\sim $ 1000 unit cells) is much greater than the out-of-plane correlation length $\xi_{c}$ $\sim $ 250 \AA ($\sim $ 20 unit cells), though we note that the latter could be limited by the approximate $1000$ \AA  thickness of the films in the $c$ direction. The non-integer out-of-plane projection of the ordering vector ($L \approx 4.84$) reveals an incommensurate nature of the ordering pattern along the \textit{c} axis.

To uncover the nature of these reflections, we have studied their evolution as a function of temperature, photon energy, and azimuthal angle.  Figure\,\ref{fig2}A shows a series of representative rocking curves at ${\mathbf{Q}}_{N} = \left( 1/2, 1/2, L \right)$ for various temperatures, with the integrated intensity extracted from Gaussian fits and plotted in Fig.\,\ref{fig2}B. A clear transition occurs at $T \approx 110$\,K, which coincides with the reported N\'{e}el temperature for polycrystalline samples revealed by bulk magnetometry (see susceptibility trace from Ref. \citep{sakurai_synthesis_2014} in Fig.\,\ref{fig2}B), as well as muon spin rotation measurements \citep{nozaki_2018,sugiyama_microscopic_2014}. In Fig.\,\ref{fig2}C, we plot the integrated intensity of the rocking curve for the $ {\mathbf{Q}}_{N} $ reflection as a function of the photon energy at $T = 20$\,K, showing a clear resonant enhancement around $E = 6.009$\,keV, at the main Cr-$K$ edge, measured using near edge X-ray absorption (XANES) spectrum measured in fluorescence mode. Given the close correspondence of the transition temperature of this peak with the reported N\'{e}el temperature, we performed \textit{ab initio} calculations using the FDMNES program to model the RXS intensity resulting from N\'{e}el-type magnetic order \citep{joly_self-consistency_2009}. The RXS intensity and XANES curves are overlaid onto the experimental data points in Fig.\,\ref{fig2}C, confirming the resonant enhancement of the magnetic peak at the Cr-$1s \rightarrow 4p$ transition (for more details, see Supplementary Information).  

To further examine the symmetry of the ordered phase, we also study the azimuthal angle dependence of the integrated intensity at $ {\mathbf{Q}}_{N} $. RXS scans across $ {\mathbf{Q}}_{N} $ were collected for different values of the azimuthal $\psi$, which corresponds to a rotation around the axis parallel to $ {\mathbf{Q}}_{N} $. Here, an azimuthal angle $\psi = 0$ coincides with a sample orientation that has the $\left( 1, 1, L \right)$ crystallographic direction lying in the scattering plane. The clear 2-fold modulation of the RXS intensity as a function of the azimuthal angle, shown in Fig.\,\ref{fig2}D, is consistent with a magnetic structure with moments oriented along $\left( 1, -1, 0 \right)$, i.e. perpendicular to the in-plane projection of the N\'{e}el ordering vector. The collective information, including the resonance energy profile, azimuthal symmetry and polarization restrictions (from the incoming photons) are most consistent with magnetic scattering. This interpretation is further bolstered by the close correspondence between the observed onset temperature and previous reports of the N\'{e}el transition \citep{sakurai_synthesis_2014,nozaki_2018,sugiyama_microscopic_2014}. We also note that, among the possible high-symmetry magnetic space subgroups, the one determined to be most consistent with the azimuthal symmetry is equivalent to that found in other isostructural compounds, such as La$_2$CuO$_4$ \cite{vaknin_1987}.

\begin{figure}[H]
\centering
\includegraphics[width=1\columnwidth]{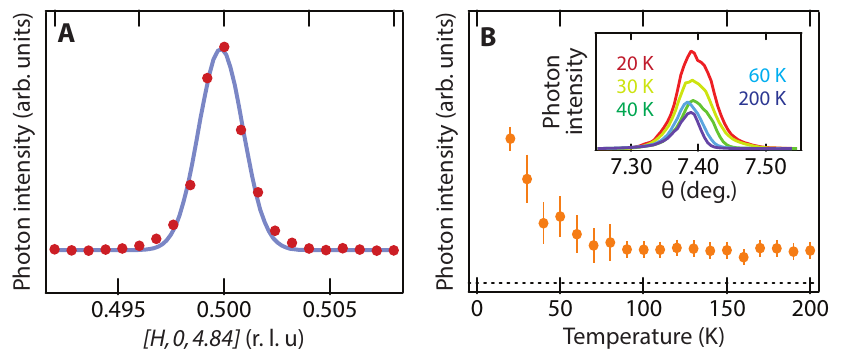}
\caption{\label{fig3} \textbf{Stripe-like electronic ordering along $\left[ H, 0 \right]$.} \textbf{A}: representative momentum scans across $ {\mathbf{Q}}_{s} = \left( 1/2, 0, L \right)$ at a photon energy $E = 6.009$\,keV. \textbf{B}: integrated RXS intensity at $ {\mathbf{Q}}_{s} $ vs. temperature; the inset shows a series of RXS rocking curves across $ {\mathbf{Q}}_{s} $ for a few representative temperatures.} 
\end{figure}
 
Having identified the nature and microscopic traits of the ordered state below $T_N \sim 110$\,K, we have focused on the $\left[ H, 0 \right]$ reciprocal space direction, to search for additional ordering instabilities. This further exploration was motivated by previous heat capacity data hinting at a second ordering transition around $140$\,K, whose nature has remained undisclosed \citep{sakurai_synthesis_2014}. Consistently with a dual-instability scenario, we have observed a resonant reflection at the wavevector for period-2 stripe order $ {\mathbf{Q}}_{s} = \left( 1/2, 0, L \right)$ (with $L = 4.84$ here, as well) as shown in Figure\,\ref{fig3}A.  Similar to the reflection at the N\'{e}el vector $ {\mathbf{Q}}_{N}$, the diffraction peak at the stripe ordering vector also exhibits a resonant enhancement as the photon energy is tuned to $E = 6.009$\,keV (see Supplementary Information).  The integrated intensity of the rocking curves as a function of temperature is shown in Fig.\,\ref{fig3}B (RXS scans at representative temperatures are reported in the inset). Surprisingly, the transition was seen to occur around $T_s \sim 50$\,K, namely at a significantly lower temperature than the proposed orbital ordering transition. Here, we should note that a remnant scattering signal persists above the transition and up to the highest measured temperature (200\,K), yet shows no variation around 140\,K. This unusual temperature dependence has been noted in other transition metal compounds with coupled spin and orbital order parameters (OPs), in particular KCuF$_3$ \cite{paolasini_2002}, though we cannot rule out additional OPs of possibly different symmetry contributing to the remnant signal above the clear transition at $T_s$. We reconcile the puzzle of this second transition at $T_s$  by noting that the ordering temperature is close to that of the appearance of a second oscillatory component near $45$\,K as recorded by zero-field muon spin rotation (ZF-$\mu^+$SR) measurements on polycrystalline samples \cite{sugiyama_microscopic_2014}.

\begin{figure}
\centering
\includegraphics[width=1\columnwidth]{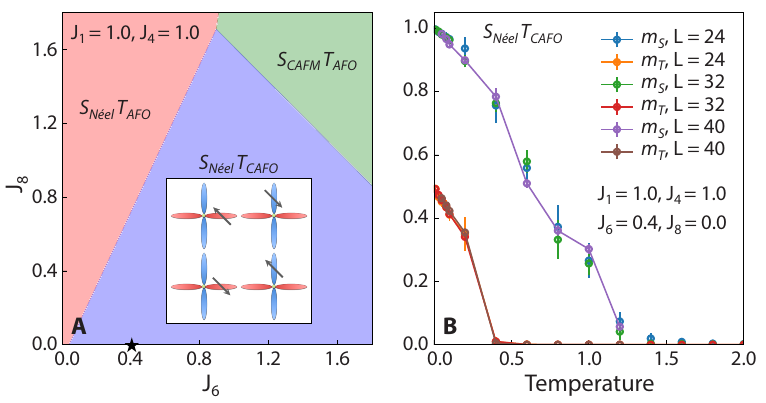}
\caption{\label{fig4} \textbf{Phase diagrams at zero and nonzero temperatures.} \textbf{A}: The
phase diagram at zero temperature for the Kugel-Khomskii model with $J_1=J_4=1.0$, based on a site-factorized wavefunction approach. Three stable phases appear in the phase diagram, including $S_{N\acute{e}el}T_{CAFO}$, $S_{N\acute{e}el}T_{AFO}$, and $S_{CAFM}T_{AFO}$. The inset is a schematic illustration of the $S_{N\acute{e}el}T_{CAFO}$ ordered state. The blue and red ovals represent $\vert xz \rangle$ and $\vert yz \rangle$ orbitals, respectively. The black arrows denote the spin state on the $\vert xz \rangle$ or $\vert yz \rangle$ orbital. S$_{N\acute{e}el}$ and S$_{CAFM}$ stand for N\'{e}el order and stripe antiferromagnetic order, respectively. $T_{CAFO}$ marks a stripe-like orbital order with momentum $(\pi,0)/(0,\pi)$,and T$_{AFO}$ represents N\'{e}el-like orbital order with momentum $(\pi,\pi)$. \textbf{B}: The temperature-dependence of the magnetic ($m_S$) and orbital ($m_T$) order parameters from classical Monte Carlo simulations of the phase $S_{N\acute{e}el}T_{CAFO}$ with $J_1=J_4=1.0,J_6=0.4,J_8=0.0$ on $L=24,32,40$ clusters (marked by the black star in panel \textbf{A}).  The model parameters $J_i$ are set as $J_i=t^2_i/U$, where $U$ is the on-site interaction and $t_i$ represent hopping amplitudes between orbitals up to next-nearest neighbors.  $J_1$ and $J_4$ are the nearest-neighbor couplings, and $J_6$ and $J_8$ the next-nearest-neighbor couplings. Further details are given in the Supplementary Information.}
\vspace{-0.5cm}
\end{figure}
 
To understand the microscopic physics behind the observed scattering peaks, we examine an effective spin-$1$ Kugel-Khomskii model for $\alpha$-Sr$_2$CrO$_4$ (see Supplementary Information for additional details). In this compound, the necessarily high-spin $3d^2$ electronic configuration form an active $t_{2g}$ orbital subspace. Of particular interest are results from density functional theory (DFT) calculations showing a reversed crystal field splitting compared to typical tetragonal systems with an elongated c-axis lattice parameter \citep{ishikawa_reversed_2017}. Within this scenario, two electrons occupy the $\left\lbrace \vert xy \rangle, \vert xz \rangle, \vert yz \rangle \right\rbrace$ $t_{2g}$ orbital manifold, with a proposed negative energy level splitting $\Delta \sim -0.6eV$ between the low-lying $\vert xy \rangle$ and $\vert xz \rangle $/$ \vert yz \rangle$ orbitals. This reversed crystal field leaves a single electron in the two-fold degenerate $\left\lbrace \vert xz \rangle, \vert yz \rangle \right\rbrace$ submanifold, setting the stage for active orbital physics as encoded in a pseudospin-1/2 degree of freedom. In order to elucidate the essential physics arising from the interplay of the spin and orbital degrees of freedom in $\alpha$-Sr$_2$CrO$_4$ \citep{oles_fingerprints_2005,oles_one-dimensional_2007,kruger_spin-orbital_2009}, we consider the spin-$1$ Kugel-Khomskii Hamiltonian as the effective model:

\begin{equation}
\small
H_{KK}^{(i,j)} = -\frac{1}{3}({\bf S}_i\cdot {\bf S}_j+2)Q^{(1)}({\bf T}_i,{\bf T}_j)+ \frac{1}{3}({\bf S}_i\cdot {\bf S}_j-1) Q^{(2)}({\bf T}_i,{\bf T}_j),
\end{equation}
with
\begin{eqnarray}\label{pspin}
&&Q^{(n)}({\bf T}_i,{\bf T}_j)=\nonumber \\
&&f_{zz}^{(n)}T_{i}^{z}T_{j}^{z}+\frac{1}{2} f_{+-}^{(n)}(T_{i}^{+}T_{j}^{-}+T_{i}^{-}T_{j}^{+})+f_{z}^{(n)}(T_{i}^{z}+T_{j}^{z})\nonumber \\
&&+\frac{1}{2} f_{++}^{(n)}(T_{i}^{+}T_{j}^{+}+T_{i}^{-}T_{j}^{-})+f_{zx}^{(n)}(T_{i}^{z}T_{j}^{x}+T_{i}^{x}T_{j}^{z})\nonumber \\
&&+f_{x}^{(n)}(T_{i}^{x}+T_{j}^{x})  +f_{0}^{(n)},
\end{eqnarray}
where ${\bf S}_i$ and ${\bf T}_i$ are the spin-$1$ and pseudospin-$1/2$ operators, respectively. $f^{(n)}$ with $n=1,2$ are functions of microscopic parameters that include the hopping amplitudes between orbitals up to next-nearest neighbor and the onsite Hubbard interactions and Hund's coupling (see Supplementary Information for details). 
We have performed calculations using a combination of mean-field and classical Monte Carlo methods.
As shown in Fig.\,\ref{fig4}A, we find S$_{N\acute{e}el}$T$_{CAFO}$, a N\'{e}el order accompanied by a stripe-like orbital order, over an extended parameter regime in the phase diagram. This phase corresponds to the same ordering vector as found in RXS experiments.  In Fig.\ref{fig4}B, we show the results of classical Monte Carlo simulations for the evolution of the magnetic and orbital order parameters vs. temperature for the phase S$_{N\acute{e}el}$T$_{CAFO}$.  The numerical results clearly show that, as temperature is lowered, the magnetic order parameter develops first, which is followed by the emergence of the stripe-like orbital order.

The proposed theoretical description captures the salient features of the experimental observations, including the ordering wavevectors ($\mathbf{Q}_s$ and $\mathbf{Q}_N$) and their respective transition temperatures. The N\'{e}el spin order setting in at the higher-temperature transition is consistent with our RXS measurements (see Supplementary Information) as well as the magnetic measurements in polycrystalline samples \cite{sakurai_synthesis_2014,sugiyama_microscopic_2014}.  In addition, because the orbital order below the lower-temperature transition develops in a time-reversal-broken background, it is expected to develop an additional orbital moment at the same wavevector, in the presence of spin-orbit coupling. Given that the involved orbitals are $\vert xz \rangle$ and $\vert yz \rangle$, the orbital moment is likely to dominantly point along the $c$-axis, and this appears to be the most consistent with the existing RXS azimuthal dependence (see Supplementary Information)  although further spectroscopic studies will be needed in order to reach a firm conclusion.  Finally, while our model study has focused on the largest part of the energetics, a complete understanding will require the addition of couplings with smaller magnitude, such as the interlayer couplings.
 
In summary, our experimental and theoretical results uncover the microscopic nature of the antiferromagnetic state in $\alpha$-Sr$_2$CrO$_4$, which is here shown using resonant X-ray scattering to arise from collinear N\'{e}el-type order at ${\mathbf{Q}}_{N} = \left( \pm 1/2, \pm 1/2, L \right)$ with the spin axis in the CrO$_2$ plane, in close analogy with the parent state of cuprate superconductors. In addition to N\'{e}el order, we have detected coexisting stripe-like order setting in at a lower temperature $T_s \sim 50$\,K, with ordering vector $ {\mathbf{Q}}_{s} = \left( 1/2, 0, L \right)$. 
A Monte Carlo study of the spin-1 Kugel-Khomskii Hamiltonian captures the observed experimental signatures of N\'{e}el and stripe phases and their ordering sequence, thereby pointing to the physical realization of this model for a system of $d^2$ ions on a square lattice.  Our study singles out $\alpha$-Sr$_2$CrO$_4$ as a unique platform to explore the rich electronic phases in a Kugel-Khomskii-like spin-orbital system, and underscores the great scientific potential of a unique transition metal oxide, whose electronic phase diagram as a function of carrier doping is yet to be explored.

\begin{acknowledgments}

We are grateful to A. Chubukov, D. Puggioni, J.M. Tranquada, and P.A. Lee for insightful discussions. Work at MIT (Z.H.Z., C.A.O., J.L., J.P., and R.C.) has been supported by the Department of Energy, Office of Science, Office of Basic Energy Sciences, under Award Number DE-SC0019126. Part of this research concerning sample synthesis (Z.H.Z.) has been supported by NSF through the Massachusetts Institute of Technology Materials Research Science and Engineering Center DMR – 1419807. Work at Rice (W.H. and Q.S.) has in part been supported by the U.S. Department of Energy, Office of Science, Basic Energy Sciences, under Award No. DE-SC0018197 and by the Robert A. Welch Foundation Grant No. C-1411. R.C. acknowledges support from the Alfred P. Sloan Foundation. M.N. was supported by the Materials Sciences and Engineering Division, Basic Energy Sciences, Office of Science, US DOE. J. P. acknowledges financial support by the Swiss National Science Foundation Early Postdoc Mobility Fellowship Project No. P2FRP2\_171824 and P400P2\_180744. This research used beamline 4-ID of the National Synchrotron Light Source II, a U.S. Department of Energy (DOE) Office of Science User Facility operated for the DOE Office of Science by Brookhaven National Laboratory under Contract No. DE-SC0012704.

\end{acknowledgments}


\begin{thebibliography}{35}
\expandafter\ifx\csname natexlab\endcsname\relax\def\natexlab#1{#1}\fi
\expandafter\ifx\csname bibnamefont\endcsname\relax
  \def\bibnamefont#1{#1}\fi
\expandafter\ifx\csname bibfnamefont\endcsname\relax
  \def\bibfnamefont#1{#1}\fi
\expandafter\ifx\csname citenamefont\endcsname\relax
  \def\citenamefont#1{#1}\fi
\expandafter\ifx\csname url\endcsname\relax
  \def\url#1{\texttt{#1}}\fi
\expandafter\ifx\csname urlprefix\endcsname\relax\def\urlprefix{URL }\fi
\providecommand{\bibinfo}[2]{#2}
\providecommand{\eprint}[2][]{\url{#2}}

\bibitem[{\citenamefont{Streltsov and Khomskii}(2017)}]{streltsov_orbital_2017}
\bibinfo{author}{\bibfnamefont{S.~V.} \bibnamefont{Streltsov}}
  \bibnamefont{and} \bibinfo{author}{\bibfnamefont{D.~I.}
  \bibnamefont{Khomskii}}, \bibinfo{journal}{Physics-Uspekhi}
  \textbf{\bibinfo{volume}{60}}, \bibinfo{pages}{1121} (\bibinfo{year}{2017}),
  ISSN \bibinfo{issn}{1063-7869},
  \urlprefix\url{https://iopscience.iop.org/article/10.3367/UFNe.2017.08.038196/meta}.

\bibitem[{\citenamefont{Tokura and Nagaosa}(2000)}]{tokura_orbital_2000}
\bibinfo{author}{\bibfnamefont{Y.}~\bibnamefont{Tokura}} \bibnamefont{and}
  \bibinfo{author}{\bibfnamefont{N.}~\bibnamefont{Nagaosa}},
  \bibinfo{journal}{Science} \textbf{\bibinfo{volume}{288}},
  \bibinfo{pages}{462} (\bibinfo{year}{2000}), ISSN \bibinfo{issn}{0036-8075,
  1095-9203},
  \urlprefix\url{https://science.sciencemag.org/content/288/5465/462}.

\bibitem[{\citenamefont{Murakami et~al.}(1998)\citenamefont{Murakami, Hill,
  Gibbs, Blume, Koyama, Tanaka, Kawata, Arima, Tokura, Hirota
  et~al.}}]{murakami_resonant_1998}
\bibinfo{author}{\bibfnamefont{Y.}~\bibnamefont{Murakami}},
  \bibinfo{author}{\bibfnamefont{J.~P.} \bibnamefont{Hill}},
  \bibinfo{author}{\bibfnamefont{D.}~\bibnamefont{Gibbs}},
  \bibinfo{author}{\bibfnamefont{M.}~\bibnamefont{Blume}},
  \bibinfo{author}{\bibfnamefont{I.}~\bibnamefont{Koyama}},
  \bibinfo{author}{\bibfnamefont{M.}~\bibnamefont{Tanaka}},
  \bibinfo{author}{\bibfnamefont{H.}~\bibnamefont{Kawata}},
  \bibinfo{author}{\bibfnamefont{T.}~\bibnamefont{Arima}},
  \bibinfo{author}{\bibfnamefont{Y.}~\bibnamefont{Tokura}},
  \bibinfo{author}{\bibfnamefont{K.}~\bibnamefont{Hirota}},
  \bibnamefont{et~al.}, \bibinfo{journal}{Physical Review Letters}
  \textbf{\bibinfo{volume}{81}}, \bibinfo{pages}{582} (\bibinfo{year}{1998}),
  \urlprefix\url{https://link.aps.org/doi/10.1103/PhysRevLett.81.582}.

\bibitem[{\citenamefont{Caciuffo et~al.}(2002)\citenamefont{Caciuffo,
  Paolasini, Sollier, Ghigna, Pavarini, van~den Brink, and
  Altarelli}}]{caciuffo_resonant_2002}
\bibinfo{author}{\bibfnamefont{R.}~\bibnamefont{Caciuffo}},
  \bibinfo{author}{\bibfnamefont{L.}~\bibnamefont{Paolasini}},
  \bibinfo{author}{\bibfnamefont{A.}~\bibnamefont{Sollier}},
  \bibinfo{author}{\bibfnamefont{P.}~\bibnamefont{Ghigna}},
  \bibinfo{author}{\bibfnamefont{E.}~\bibnamefont{Pavarini}},
  \bibinfo{author}{\bibfnamefont{J.}~\bibnamefont{van~den Brink}},
  \bibnamefont{and}
  \bibinfo{author}{\bibfnamefont{M.}~\bibnamefont{Altarelli}},
  \bibinfo{journal}{Physical Review B} \textbf{\bibinfo{volume}{65}},
  \bibinfo{pages}{174425} (\bibinfo{year}{2002}),
  \urlprefix\url{https://link.aps.org/doi/10.1103/PhysRevB.65.174425}.

\bibitem[{\citenamefont{Noguchi et~al.}(2000)\citenamefont{Noguchi, Nakazawa,
  Oka, Arima, Wakabayashi, Nakao, and Murakami}}]{noguchi_synchrotron_2000}
\bibinfo{author}{\bibfnamefont{M.}~\bibnamefont{Noguchi}},
  \bibinfo{author}{\bibfnamefont{A.}~\bibnamefont{Nakazawa}},
  \bibinfo{author}{\bibfnamefont{S.}~\bibnamefont{Oka}},
  \bibinfo{author}{\bibfnamefont{T.}~\bibnamefont{Arima}},
  \bibinfo{author}{\bibfnamefont{Y.}~\bibnamefont{Wakabayashi}},
  \bibinfo{author}{\bibfnamefont{H.}~\bibnamefont{Nakao}}, \bibnamefont{and}
  \bibinfo{author}{\bibfnamefont{Y.}~\bibnamefont{Murakami}},
  \bibinfo{journal}{Physical Review B} \textbf{\bibinfo{volume}{62}},
  \bibinfo{pages}{R9271} (\bibinfo{year}{2000}),
  \urlprefix\url{https://link.aps.org/doi/10.1103/PhysRevB.62.R9271}.

\bibitem[{\citenamefont{Yi et~al.}(2017)\citenamefont{Yi, Zhang, Shen, and
  Lu}}]{yi_role_2017}
\bibinfo{author}{\bibfnamefont{M.}~\bibnamefont{Yi}},
  \bibinfo{author}{\bibfnamefont{Y.}~\bibnamefont{Zhang}},
  \bibinfo{author}{\bibfnamefont{Z.-X.} \bibnamefont{Shen}}, \bibnamefont{and}
  \bibinfo{author}{\bibfnamefont{D.}~\bibnamefont{Lu}}, \bibinfo{journal}{npj
  Quantum Materials} \textbf{\bibinfo{volume}{2}}, \bibinfo{pages}{57}
  (\bibinfo{year}{2017}), ISSN \bibinfo{issn}{2397-4648},
  \urlprefix\url{https://www.nature.com/articles/s41535-017-0059-y}.

\bibitem[{\citenamefont{Liu et~al.}(2010)\citenamefont{Liu, Hu, Qian, Fobes,
  Mao, Bao, Reehuis, Kimber, Proke\v{s}, Matas et~al.}}]{liu_0_2010}
\bibinfo{author}{\bibfnamefont{T.~J.} \bibnamefont{Liu}},
  \bibinfo{author}{\bibfnamefont{J.}~\bibnamefont{Hu}},
  \bibinfo{author}{\bibfnamefont{B.}~\bibnamefont{Qian}},
  \bibinfo{author}{\bibfnamefont{D.}~\bibnamefont{Fobes}},
  \bibinfo{author}{\bibfnamefont{Z.~Q.} \bibnamefont{Mao}},
  \bibinfo{author}{\bibfnamefont{W.}~\bibnamefont{Bao}},
  \bibinfo{author}{\bibfnamefont{M.}~\bibnamefont{Reehuis}},
  \bibinfo{author}{\bibfnamefont{S.~A.~J.} \bibnamefont{Kimber}},
  \bibinfo{author}{\bibfnamefont{K.}~\bibnamefont{Proke\v{s}}},
  \bibinfo{author}{\bibfnamefont{S.}~\bibnamefont{Matas}},
  \bibnamefont{et~al.}, \bibinfo{journal}{Nature Materials}
  \textbf{\bibinfo{volume}{9}}, \bibinfo{pages}{718} (\bibinfo{year}{2010}),
  ISSN \bibinfo{issn}{1476-4660},
  \urlprefix\url{https://www.nature.com/articles/nmat2800}.

\bibitem[{\citenamefont{Stewart}(2011)}]{stewart_superconductivity_2011}
\bibinfo{author}{\bibfnamefont{G.~R.} \bibnamefont{Stewart}},
  \bibinfo{journal}{Reviews of Modern Physics} \textbf{\bibinfo{volume}{83}},
  \bibinfo{pages}{1589} (\bibinfo{year}{2011}),
  \urlprefix\url{https://link.aps.org/doi/10.1103/RevModPhys.83.1589}.

\bibitem[{\citenamefont{Lee et~al.}(2006)\citenamefont{Lee, Nagaosa, and
  Wen}}]{lee_doping_2006}
\bibinfo{author}{\bibfnamefont{P.~A.} \bibnamefont{Lee}},
  \bibinfo{author}{\bibfnamefont{N.}~\bibnamefont{Nagaosa}}, \bibnamefont{and}
  \bibinfo{author}{\bibfnamefont{X.-G.} \bibnamefont{Wen}},
  \bibinfo{journal}{Reviews of Modern Physics} \textbf{\bibinfo{volume}{78}},
  \bibinfo{pages}{17} (\bibinfo{year}{2006}),
  \urlprefix\url{https://link.aps.org/doi/10.1103/RevModPhys.78.17}.

\bibitem[{\citenamefont{Chamberland}(1967)}]{chamberland_preparation_1967}
\bibinfo{author}{\bibfnamefont{B.~L.} \bibnamefont{Chamberland}},
  \bibinfo{journal}{Solid State Communications} \textbf{\bibinfo{volume}{5}},
  \bibinfo{pages}{663} (\bibinfo{year}{1967}), ISSN \bibinfo{issn}{0038-1098},
  \urlprefix\url{http://www.sciencedirect.com/science/article/pii/0038109867900889}.

\bibitem[{\citenamefont{Chamberland and
  Moeller}(1972)}]{chamberland_study_1972}
\bibinfo{author}{\bibfnamefont{B.~L.} \bibnamefont{Chamberland}}
  \bibnamefont{and} \bibinfo{author}{\bibfnamefont{C.~W.}
  \bibnamefont{Moeller}}, \bibinfo{journal}{Journal of Solid State Chemistry}
  \textbf{\bibinfo{volume}{5}}, \bibinfo{pages}{39} (\bibinfo{year}{1972}),
  ISSN \bibinfo{issn}{0022-4596},
  \urlprefix\url{http://www.sciencedirect.com/science/article/pii/0022459672900060}.

\bibitem[{\citenamefont{Goodenough et~al.}(1968)\citenamefont{Goodenough,
  Longo, and Kafalas}}]{goodenough_band_1968}
\bibinfo{author}{\bibfnamefont{J.~B.} \bibnamefont{Goodenough}},
  \bibinfo{author}{\bibfnamefont{J.~M.} \bibnamefont{Longo}}, \bibnamefont{and}
  \bibinfo{author}{\bibfnamefont{J.~A.} \bibnamefont{Kafalas}},
  \bibinfo{journal}{Materials Research Bulletin} \textbf{\bibinfo{volume}{3}},
  \bibinfo{pages}{471} (\bibinfo{year}{1968}), ISSN \bibinfo{issn}{0025-5408},
  \urlprefix\url{http://www.sciencedirect.com/science/article/pii/0025540868900706}.

\bibitem[{\citenamefont{Zhou et~al.}(2006)\citenamefont{Zhou, Jin, Long, Yang,
  and Goodenough}}]{zhou_anomalous_2006}
\bibinfo{author}{\bibfnamefont{J.-S.} \bibnamefont{Zhou}},
  \bibinfo{author}{\bibfnamefont{C.-Q.} \bibnamefont{Jin}},
  \bibinfo{author}{\bibfnamefont{Y.-W.} \bibnamefont{Long}},
  \bibinfo{author}{\bibfnamefont{L.-X.} \bibnamefont{Yang}}, \bibnamefont{and}
  \bibinfo{author}{\bibfnamefont{J.~B.} \bibnamefont{Goodenough}},
  \bibinfo{journal}{Physical Review Letters} \textbf{\bibinfo{volume}{96}},
  \bibinfo{pages}{046408} (\bibinfo{year}{2006}),
  \urlprefix\url{https://link.aps.org/doi/10.1103/PhysRevLett.96.046408}.

\bibitem[{\citenamefont{Ortega-San-Martin
  et~al.}(2007)\citenamefont{Ortega-San-Martin, Williams, Rodgers, Attfield,
  Heymann, and Huppertz}}]{ortega-san-martin_microstrain_2007}
\bibinfo{author}{\bibfnamefont{L.}~\bibnamefont{Ortega-San-Martin}},
  \bibinfo{author}{\bibfnamefont{A.~J.} \bibnamefont{Williams}},
  \bibinfo{author}{\bibfnamefont{J.}~\bibnamefont{Rodgers}},
  \bibinfo{author}{\bibfnamefont{J.~P.} \bibnamefont{Attfield}},
  \bibinfo{author}{\bibfnamefont{G.}~\bibnamefont{Heymann}}, \bibnamefont{and}
  \bibinfo{author}{\bibfnamefont{H.}~\bibnamefont{Huppertz}},
  \bibinfo{journal}{Physical Review Letters} \textbf{\bibinfo{volume}{99}},
  \bibinfo{pages}{255701} (\bibinfo{year}{2007}),
  \urlprefix\url{https://link.aps.org/doi/10.1103/PhysRevLett.99.255701}.

\bibitem[{\citenamefont{Komarek et~al.}(2008)\citenamefont{Komarek, Streltsov,
  Isobe, Möller, Hoelzel, Senyshyn, Trots, Fern\'{a}ndez-Díaz, Hansen, Gotou
  et~al.}}]{komarek_2008}
\bibinfo{author}{\bibfnamefont{A.~C.} \bibnamefont{Komarek}},
  \bibinfo{author}{\bibfnamefont{S.~V.} \bibnamefont{Streltsov}},
  \bibinfo{author}{\bibfnamefont{M.}~\bibnamefont{Isobe}},
  \bibinfo{author}{\bibfnamefont{T.}~\bibnamefont{M\"{o}ller}},
  \bibinfo{author}{\bibfnamefont{M.}~\bibnamefont{Hoelzel}},
  \bibinfo{author}{\bibfnamefont{A.}~\bibnamefont{Senyshyn}},
  \bibinfo{author}{\bibfnamefont{D.}~\bibnamefont{Trots}},
  \bibinfo{author}{\bibfnamefont{M.~T.} \bibnamefont{Fern\'{a}ndez-D\'{i}az}},
  \bibinfo{author}{\bibfnamefont{T.}~\bibnamefont{Hansen}},
  \bibinfo{author}{\bibfnamefont{H.}~\bibnamefont{Gotou}},
  \bibinfo{author}{\bibfnamefont{T.}~\bibnamefont{Yagi}},
  \bibinfo{author}{\bibfnamefont{Y.}~\bibnamefont{Ueda}},
  \bibinfo{author}{\bibfnamefont{V.~I.} \bibnamefont{Anisimov}},
  \bibinfo{author}{\bibfnamefont{M.}~\bibnamefont{Gruninger}},
  \bibinfo{author}{\bibfnamefont{D.~I.}~\bibnamefont{Khomskii}},
  \bibinfo{author}{\bibfnamefont{M.}~\bibnamefont{Braden}},
  \bibinfo{journal}{Physical Review Letters}
  \textbf{\bibinfo{volume}{101}}, \bibinfo{pages}{167204}
  (\bibinfo{year}{2008}),
  \urlprefix\url{https://link.aps.org/doi/10.1103/PhysRevLett.101.167204}.

\bibitem[{\citenamefont{Zhu et~al.}(2013)\citenamefont{Zhu, Rueckert, Budnick,
  Hines, Jain, Zhang, and Wells}}]{zhu_magnetic_2013}
\bibinfo{author}{\bibfnamefont{Z.~H.} \bibnamefont{Zhu}},
  \bibinfo{author}{\bibfnamefont{F.~J.} \bibnamefont{Rueckert}},
  \bibinfo{author}{\bibfnamefont{J.~I.} \bibnamefont{Budnick}},
  \bibinfo{author}{\bibfnamefont{W.~A.} \bibnamefont{Hines}},
  \bibinfo{author}{\bibfnamefont{M.}~\bibnamefont{Jain}},
  \bibinfo{author}{\bibfnamefont{H.}~\bibnamefont{Zhang}}, \bibnamefont{and}
  \bibinfo{author}{\bibfnamefont{B.~O.} \bibnamefont{Wells}},
  \bibinfo{journal}{Physical Review B} \textbf{\bibinfo{volume}{87}},
  \bibinfo{pages}{195129} (\bibinfo{year}{2013}),
  \urlprefix\url{https://link.aps.org/doi/10.1103/PhysRevB.87.195129}.

\bibitem[{\citenamefont{Ar\'{e}valo-L\'{o}pez and
  Paul~Attfield}(2015)}]{arevalo-lopez_high-pressure_2015}
\bibinfo{author}{\bibfnamefont{A.~M.} \bibnamefont{Ar\'{e}valo-L\'{o}pez}}
  \bibnamefont{and}
  \bibinfo{author}{\bibfnamefont{J.}~\bibnamefont{Paul~Attfield}},
  \bibinfo{journal}{Journal of Solid State Chemistry}
  \textbf{\bibinfo{volume}{232}}, \bibinfo{pages}{236} (\bibinfo{year}{2015}),
  ISSN \bibinfo{issn}{0022-4596},
  \urlprefix\url{http://www.sciencedirect.com/science/article/pii/S002245961530181X}.

\bibitem[{\citenamefont{Giovannetti et~al.}(2014)\citenamefont{Giovannetti,
  Aichhorn, and Capone}}]{giovannetti_cooperative_2014}
\bibinfo{author}{\bibfnamefont{G.}~\bibnamefont{Giovannetti}},
  \bibinfo{author}{\bibfnamefont{M.}~\bibnamefont{Aichhorn}}, \bibnamefont{and}
  \bibinfo{author}{\bibfnamefont{M.}~\bibnamefont{Capone}},
  \bibinfo{journal}{Physical Review B} \textbf{\bibinfo{volume}{90}},
  \bibinfo{pages}{245134} (\bibinfo{year}{2014}),
  \urlprefix\url{https://link.aps.org/doi/10.1103/PhysRevB.90.245134}.

\bibitem[{\citenamefont{Jin et~al.}(2014)\citenamefont{Jin, Ahn, Jung, and
  Lee}}]{jin_strain_2014}
\bibinfo{author}{\bibfnamefont{H.-S.} \bibnamefont{Jin}},
  \bibinfo{author}{\bibfnamefont{K.-H.} \bibnamefont{Ahn}},
  \bibinfo{author}{\bibfnamefont{M.-C.} \bibnamefont{Jung}}, \bibnamefont{and}
  \bibinfo{author}{\bibfnamefont{K.-W.} \bibnamefont{Lee}},
  \bibinfo{journal}{Physical Review B} \textbf{\bibinfo{volume}{90}},
  \bibinfo{pages}{205124} (\bibinfo{year}{2014}),
  \urlprefix\url{https://link.aps.org/doi/10.1103/PhysRevB.90.205124}.

\bibitem[{\citenamefont{Jeanneau et~al.}(2017)\citenamefont{Jeanneau,
  Toulemonde, Remenyi, Sulpice, Colin, Nassif, Suard, Salas~Colera, Castro, Gay
  et~al.}}]{jeanneau_singlet_2017}
\bibinfo{author}{\bibfnamefont{J.}~\bibnamefont{Jeanneau}},
  \bibinfo{author}{\bibfnamefont{P.}~\bibnamefont{Toulemonde}},
  \bibinfo{author}{\bibfnamefont{G.}~\bibnamefont{Remenyi}},
  \bibinfo{author}{\bibfnamefont{A.}~\bibnamefont{Sulpice}},
  \bibinfo{author}{\bibfnamefont{C.}~\bibnamefont{Colin}},
  \bibinfo{author}{\bibfnamefont{V.}~\bibnamefont{Nassif}},
  \bibinfo{author}{\bibfnamefont{E.}~\bibnamefont{Suard}},
  \bibinfo{author}{\bibfnamefont{E.}~\bibnamefont{Salas~Colera}},
  \bibinfo{author}{\bibfnamefont{G.~R.} \bibnamefont{Castro}},
  \bibinfo{author}{\bibfnamefont{F.}~\bibnamefont{Gay}}, \bibnamefont{et~al.},
  \bibinfo{journal}{Physical Review Letters} \textbf{\bibinfo{volume}{118}},
  \bibinfo{pages}{207207} (\bibinfo{year}{2017}),
  \urlprefix\url{https://link.aps.org/doi/10.1103/PhysRevLett.118.207207}.

\bibitem[{\citenamefont{Ishikawa et~al.}(2017)\citenamefont{Ishikawa, Toriyama,
  Konishi, Sakurai, and Ohta}}]{ishikawa_reversed_2017}
\bibinfo{author}{\bibfnamefont{T.}~\bibnamefont{Ishikawa}},
  \bibinfo{author}{\bibfnamefont{T.}~\bibnamefont{Toriyama}},
  \bibinfo{author}{\bibfnamefont{T.}~\bibnamefont{Konishi}},
  \bibinfo{author}{\bibfnamefont{H.}~\bibnamefont{Sakurai}}, \bibnamefont{and}
  \bibinfo{author}{\bibfnamefont{Y.}~\bibnamefont{Ohta}},
  \bibinfo{journal}{Journal of the Physical Society of Japan}
  \textbf{\bibinfo{volume}{86}}, \bibinfo{pages}{033701}
  (\bibinfo{year}{2017}), ISSN \bibinfo{issn}{0031-9015},
  \urlprefix\url{https://journals.jps.jp/doi/full/10.7566/JPSJ.86.033701}.

\bibitem[{\citenamefont{Sakurai}(2014)}]{sakurai_synthesis_2014}
\bibinfo{author}{\bibfnamefont{H.}~\bibnamefont{Sakurai}},
  \bibinfo{journal}{Journal of the Physical Society of Japan}
  \textbf{\bibinfo{volume}{83}}, \bibinfo{pages}{123701}
  (\bibinfo{year}{2014}), ISSN \bibinfo{issn}{0031-9015},
  \urlprefix\url{https://journals.jps.jp/doi/abs/10.7566/JPSJ.83.123701}.

\bibitem[{\citenamefont{Nozaki et~al.}(2018)\citenamefont{Nozaki, Sakurai,
  Umegaki, Ansaldo, Morris, Hitti, Arseneau, Andreica, Amato, M{\aa}nsson
  et~al.}}]{nozaki_2018}
\bibinfo{author}{\bibfnamefont{H.}~\bibnamefont{Nozaki}},
  \bibinfo{author}{\bibfnamefont{H.}~\bibnamefont{Sakurai}},
  \bibinfo{author}{\bibfnamefont{I.}~\bibnamefont{Umegaki}},
  \bibinfo{author}{\bibfnamefont{E.~J.} \bibnamefont{Ansaldo}},
  \bibinfo{author}{\bibfnamefont{G.~D.} \bibnamefont{Morris}},
  \bibinfo{author}{\bibfnamefont{B.}~\bibnamefont{Hitti}},
  \bibinfo{author}{\bibfnamefont{D.~J.} \bibnamefont{Arseneau}},
  \bibinfo{author}{\bibfnamefont{D.}~\bibnamefont{Andreica}},
  \bibinfo{author}{\bibfnamefont{A.}~\bibnamefont{Amato}},
  \bibinfo{author}{\bibfnamefont{M.}~\bibnamefont{M{\aa}nsson}},
  \bibnamefont{et~al.} (\bibinfo{year}{2018}), vol.~\bibinfo{volume}{21} of
  \emph{\bibinfo{series}{{JPS} {Conference} {Proceedings}}},
  \urlprefix\url{https://journals.jps.jp/doi/abs/10.7566/JPSCP.21.011005}.

\bibitem[{\citenamefont{Sugiyama et~al.}(2014)\citenamefont{Sugiyama, Nozaki,
  Umegaki, Higemoto, Ansaldo, Brewer, Sakurai, Kao, Yang, and
  M{\aa}unsson}}]{sugiyama_microscopic_2014}
\bibinfo{author}{\bibfnamefont{J.}~\bibnamefont{Sugiyama}},
  \bibinfo{author}{\bibfnamefont{H.}~\bibnamefont{Nozaki}},
  \bibinfo{author}{\bibfnamefont{I.}~\bibnamefont{Umegaki}},
  \bibinfo{author}{\bibfnamefont{W.}~\bibnamefont{Higemoto}},
  \bibinfo{author}{\bibfnamefont{E.~J.} \bibnamefont{Ansaldo}},
  \bibinfo{author}{\bibfnamefont{J.~H.} \bibnamefont{Brewer}},
  \bibinfo{author}{\bibfnamefont{H.}~\bibnamefont{Sakurai}},
  \bibinfo{author}{\bibfnamefont{T.-H.} \bibnamefont{Kao}},
  \bibinfo{author}{\bibfnamefont{H.-D.} \bibnamefont{Yang}}, \bibnamefont{and}
  \bibinfo{author}{\bibfnamefont{M.}~\bibnamefont{M{\aa}unsson}},
  \bibinfo{journal}{Journal of Physics: Conference Series}
  \textbf{\bibinfo{volume}{551}}, \bibinfo{pages}{012011}
  (\bibinfo{year}{2014}), ISSN \bibinfo{issn}{1742-6596},
  \urlprefix\url{http://stacks.iop.org/1742-6596/551/i=1/a=012011}.

\bibitem[{\citenamefont{Baikie et~al.}(2007)\citenamefont{Baikie, Ahmad,
  Srinivasan, Maignan, Pramana, and White}}]{baikie_crystallographic_2007}
\bibinfo{author}{\bibfnamefont{T.}~\bibnamefont{Baikie}},
  \bibinfo{author}{\bibfnamefont{Z.}~\bibnamefont{Ahmad}},
  \bibinfo{author}{\bibfnamefont{M.}~\bibnamefont{Srinivasan}},
  \bibinfo{author}{\bibfnamefont{A.}~\bibnamefont{Maignan}},
  \bibinfo{author}{\bibfnamefont{S.~S.} \bibnamefont{Pramana}},
  \bibnamefont{and} \bibinfo{author}{\bibfnamefont{T.~J.} \bibnamefont{White}},
  \bibinfo{journal}{Journal of Solid State Chemistry}
  \textbf{\bibinfo{volume}{180}}, \bibinfo{pages}{1538} (\bibinfo{year}{2007}),
  ISSN \bibinfo{issn}{0022-4596},
  \urlprefix\url{http://www.sciencedirect.com/science/article/pii/S0022459607000977}.

\bibitem[{\citenamefont{Chamberland et~al.}(1985)\citenamefont{Chamberland,
  Herrero-Fernandez, and Hewston}}]{chamberland_magnetic_1985}
\bibinfo{author}{\bibfnamefont{B.~L.} \bibnamefont{Chamberland}},
  \bibinfo{author}{\bibfnamefont{M.~P.} \bibnamefont{Herrero-Fernandez}},
  \bibnamefont{and} \bibinfo{author}{\bibfnamefont{T.~A.}
  \bibnamefont{Hewston}}, \bibinfo{journal}{Journal of Solid State Chemistry}
  \textbf{\bibinfo{volume}{59}}, \bibinfo{pages}{111} (\bibinfo{year}{1985}),
  ISSN \bibinfo{issn}{0022-4596},
  \urlprefix\url{http://www.sciencedirect.com/science/article/pii/0022459685903573}.

\bibitem[{\citenamefont{Matsuno et~al.}(2005)\citenamefont{Matsuno, Okimoto,
  Kawasaki, and Tokura}}]{matsuno_variation_2005}
\bibinfo{author}{\bibfnamefont{J.}~\bibnamefont{Matsuno}},
  \bibinfo{author}{\bibfnamefont{Y.}~\bibnamefont{Okimoto}},
  \bibinfo{author}{\bibfnamefont{M.}~\bibnamefont{Kawasaki}}, \bibnamefont{and}
  \bibinfo{author}{\bibfnamefont{Y.}~\bibnamefont{Tokura}},
  \bibinfo{journal}{Physical Review Letters} \textbf{\bibinfo{volume}{95}},
  \bibinfo{pages}{176404} (\bibinfo{year}{2005}),
  \urlprefix\url{https://link.aps.org/doi/10.1103/PhysRevLett.95.176404}.

\bibitem[{\citenamefont{Paolasini}(2014)}]{paolasini_resonant_2014}
\bibinfo{author}{\bibfnamefont{L.}~\bibnamefont{Paolasini}},
  \bibinfo{journal}{\'Ecole th\'ematique de la Soci\'et\'e Fran\c{c}aise de la
  Neutronique} \textbf{\bibinfo{volume}{13}}, \bibinfo{pages}{03002}
  (\bibinfo{year}{2014}), ISSN \bibinfo{issn}{2107-7223, 2107-7231},
  \urlprefix\url{https://www.neutron-sciences.org/articles/sfn/abs/2014/01/sfn201403002/sfn201403002.html}.

\bibitem[{\citenamefont{Matteo}(2012)}]{matteo_resonant_2012}
\bibinfo{author}{\bibfnamefont{S.~D.} \bibnamefont{Matteo}},
  \bibinfo{journal}{Journal of Physics D: Applied Physics}
  \textbf{\bibinfo{volume}{45}}, \bibinfo{pages}{163001}
  (\bibinfo{year}{2012}), ISSN \bibinfo{issn}{0022-3727},
  \urlprefix\url{https://doi.org/10.1088%2F0022-3727%2F45%2F16%2F163001}.

\bibitem[{\citenamefont{Joly et~al.}(2009)\citenamefont{Joly, Bun\v{a}u,
  Lorenzo, Gal\'era, Grenier, and Thompson}}]{joly_self-consistency_2009}
\bibinfo{author}{\bibfnamefont{Y.}~\bibnamefont{Joly}},
  \bibinfo{author}{\bibfnamefont{O.}~\bibnamefont{Bun\v{a}u}},
  \bibinfo{author}{\bibfnamefont{J.~E.} \bibnamefont{Lorenzo}},
  \bibinfo{author}{\bibfnamefont{R.~M.} \bibnamefont{Gal\'era}},
  \bibinfo{author}{\bibfnamefont{S.}~\bibnamefont{Grenier}}, \bibnamefont{and}
  \bibinfo{author}{\bibfnamefont{B.}~\bibnamefont{Thompson}},
  \bibinfo{journal}{Journal of Physics: Conference Series}
  \textbf{\bibinfo{volume}{190}}, \bibinfo{pages}{012007}
  (\bibinfo{year}{2009}), ISSN \bibinfo{issn}{1742-6596},
  \urlprefix\url{https://doi.org/10.1088%2F1742-6596%2F190%2F1%2F012007}.

\bibitem[{\citenamefont{Vaknin et~al.}(1987)\citenamefont{Vaknin, Sinha,
  Moncton, Johnston, Newsam, Safinya, and King~Jr.}}]{vaknin_1987}
\bibinfo{author}{\bibfnamefont{D.}~\bibnamefont{Vaknin}},
  \bibinfo{author}{\bibfnamefont{S.~K.} \bibnamefont{Sinha}},
  \bibinfo{author}{\bibfnamefont{D.~E.} \bibnamefont{Moncton}},
  \bibinfo{author}{\bibfnamefont{D.~C.} \bibnamefont{Johnston}},
  \bibinfo{author}{\bibfnamefont{J.~M.} \bibnamefont{Newsam}},
  \bibinfo{author}{\bibfnamefont{C.~R.} \bibnamefont{Safinya}},
  \bibnamefont{and} \bibinfo{author}{\bibfnamefont{H.~E.}
  \bibnamefont{King~Jr.}}, \bibinfo{journal}{Physical Review Letters}
  \textbf{\bibinfo{volume}{58}}, \bibinfo{pages}{2802} (\bibinfo{year}{1987}).

\bibitem[{\citenamefont{Paolasini et~al.}(2002)\citenamefont{Paolasini,
  Caciuffo, Sollier, Ghigna, and Altarelli}}]{paolasini_2002}
\bibinfo{author}{\bibfnamefont{L.}~\bibnamefont{Paolasini}},
  \bibinfo{author}{\bibfnamefont{R.}~\bibnamefont{Caciuffo}},
  \bibinfo{author}{\bibfnamefont{A.}~\bibnamefont{Sollier}},
  \bibinfo{author}{\bibfnamefont{P.}~\bibnamefont{Ghigna}}, \bibnamefont{and}
  \bibinfo{author}{\bibfnamefont{M.}~\bibnamefont{Altarelli}},
  \bibinfo{journal}{Physical Review Letters} \textbf{\bibinfo{volume}{88}},
  \bibinfo{pages}{106403} (\bibinfo{year}{2002}).

\bibitem[{\citenamefont{Ole\'{s} et~al.}(2005)\citenamefont{Ole\'{s},
  Khaliullin, Horsch, and Feiner}}]{oles_fingerprints_2005}
\bibinfo{author}{\bibfnamefont{A.~M.} \bibnamefont{Ole\'{s}}},
  \bibinfo{author}{\bibfnamefont{G.}~\bibnamefont{Khaliullin}},
  \bibinfo{author}{\bibfnamefont{P.}~\bibnamefont{Horsch}}, \bibnamefont{and}
  \bibinfo{author}{\bibfnamefont{L.~F.} \bibnamefont{Feiner}},
  \bibinfo{journal}{Physical Review B} \textbf{\bibinfo{volume}{72}},
  \bibinfo{pages}{214431} (\bibinfo{year}{2005}),
  \urlprefix\url{https://link.aps.org/doi/10.1103/PhysRevB.72.214431}.

\bibitem[{\citenamefont{Ole\'{s} et~al.}(2007)\citenamefont{Ole\'{s}, Horsch,
  and Khaliullin}}]{oles_one-dimensional_2007}
\bibinfo{author}{\bibfnamefont{A.~M.} \bibnamefont{Ole\'{s}}},
  \bibinfo{author}{\bibfnamefont{P.}~\bibnamefont{Horsch}}, \bibnamefont{and}
  \bibinfo{author}{\bibfnamefont{G.}~\bibnamefont{Khaliullin}},
  \bibinfo{journal}{Physical Review B} \textbf{\bibinfo{volume}{75}},
  \bibinfo{pages}{184434} (\bibinfo{year}{2007}),
  \urlprefix\url{https://link.aps.org/doi/10.1103/PhysRevB.75.184434}.

\bibitem[{\citenamefont{Kr\"uger et~al.}(2009)\citenamefont{Kr\"uger, Kumar,
  Zaanen, and van~den Brink}}]{kruger_spin-orbital_2009}
\bibinfo{author}{\bibfnamefont{F.}~\bibnamefont{Kr\"uger}},
  \bibinfo{author}{\bibfnamefont{S.}~\bibnamefont{Kumar}},
  \bibinfo{author}{\bibfnamefont{J.}~\bibnamefont{Zaanen}}, \bibnamefont{and}
  \bibinfo{author}{\bibfnamefont{J.}~\bibnamefont{van~den Brink}},
  \bibinfo{journal}{Physical Review B} \textbf{\bibinfo{volume}{79}},
  \bibinfo{pages}{054504} (\bibinfo{year}{2009}),
  \urlprefix\url{https://link.aps.org/doi/10.1103/PhysRevB.79.054504}.

\end{thebibliography}
\end{document}